\documentclass[twocolumn,showpacs,amsmath,amssymb,prb]{revtex4}

\usepackage{amsmath}
\usepackage{amssymb}
\usepackage{graphicx}
\usepackage{dcolumn}
\usepackage{bm}

\def\be{\begin{eqnarray}}
\def\ee{\end{eqnarray}}
\def\ba{\begin{array}}
\def\ea{\end{array}}
\def\nn{\nonumber}

\begin{document}


\title{The effect of sample properties on the electron velocity in quantum Hall bars}

\author{D. Eksi} %
\author{E. Cicek} %
\author{ A. I. Mese} %
\author{ S. Aktas} %
\address{Trakya University, Faculty of Arts and Sciences, Department of Physics, 22030 Edirne, Turkey}
\author{ A. Siddiki} %
\address{Physics Department, Arnold Sommerfeld Center for
Theoretical Physics, and
Center for NanoScience, \\
Ludwig-Maximilans-Universit\"at, Theresienstrasse 37, 80333
Munich, Germany}
\author{T. Hakio\u{g}lu}
\address{Department of Physics and National Center for Nanotechnology Research, Bilkent
University, 06690 Ankara, Turkey}%


\begin{abstract}
We report on our theoretical investigation of the effects of the
confining potential profile and sample size on the electron
velocity distribution in (narrow) quantum-Hall systems. The
electrostatic properties of the electron system are obtained by
the Thomas-Fermi-Poisson nonlinear screening theory. The electron
velocity distribution as a function of the lateral coordinate is
obtained from the slope of the screened potential at the Fermi
level and within the incompressible strips (ISs). We compare our
findings with the recent experiments.
\end{abstract}
\maketitle

\section{\label{sec:1} Introduction}
In the early electro-optical measurements performed on the two
dimensional electron systems (2DESs), the electrostatic potential
across the Hall bar was shown to exhibit local dips as a function
of the lateral coordinate across the
sample~\cite{Haren95:1198,Lorke93:1054,Ploog96:289}. The positions
of these local potential variations are strongly dependent on the
applied perpendicular magnetic field. They coincide with the
positions of the strips with finite width corresponding to integer
local filling factors where the longitudinal conductance vanishes,
i.e. $\sigma_{\ell}(x)=\sigma_{\rm{xx}}(x)=\sigma_{\rm{yy}}(x)=0$
. It was concluded that the ``expected'' quasi one dimensional
edge states can be as large as hundreds of micrometers, where
$\sigma_{\ell}(x)\neq0$. It was also shown that the current can
flow from the bulk in the magnetic field interval where the Hall
resistance does not assume its quantized value. In this regime,
however, the outermost edge states are reported to be
``invisible''~\cite{Ploog96:289}. On the other hand, the local
probe of the electrostatic potential and the longitudinal
resistivity
measurements~\cite{Ahlswede01:562,Ahlswede02:165,Yacoby00:3133,Yacoby04:328}
using scanning force microscopy and single-electron transistor
(SET) has indicated that the current is confined within finite
regions across the sample. These regions were later suggested to
be the ``incompressible'' regions, namely regions of integer local
Landau filling factors which are distributed in an inhomogeneous
manner over the sample due to the electronic nonlinear screening
as well as the boundary effects as previously
predicted~\cite{Chklovskii92:4026,Siddiki03:125315}. These
experiments are well explained by the recent theoretical
works~\cite{Guven03:115327,siddiki2004,Siddiki04:condmat}, which
take into account interaction effects by exploiting the smooth
confining potential profile within the Thomas-Fermi approximation
and also incorporating a local version of the Ohm's conductivity
model. These models contribute not only to the understanding of
the induced electric field and current distribution, but also to
the high precision nature of the low-temperature integer quantized
Hall (QH) plateaus in narrow Hall bars as a function of the
continuous lateral sample coordinate.

Recently, the edge profile a InP/InGaAs Hall sample was probed in
the ``surface photo-voltage (SPV) spectroscopy'' measurements and
it was found that the electron velocity at the edges increases
with increasing magnetic field $(B)$ respecting a square root
behavior~\cite{Karmakar05:282}, i.e. $v_{\rm{el}}(B) \propto
B^{1/2}$. In the interpretation of the data these authors used a
model in which the electrostatic bending of the Landau levels
(resulting from the confinement potential) was not taken into
account, the electron Hartree potential was neglected and instead,
the velocity distribution was modelled by an homogeneous induced
electric field. It was stated therein that, these measurements
should be re-interpreted in the light of a ``screening model''. It
is the aim of the present paper to show that their original
interpretation cannot be corrected even by including a linear
screening model (see Sec.~\ref{sec:dd}). In parallel to these
developments, the nonlinear screening was also promoted by the
importance of the recent electronic-Mach-Zehnder interferometer
experiments~\cite{Heiblum03:415,Neder06:016804} where the role of
the electron-electron interaction on the $B$ field dependence of
the edge fields was emphasized. In these latter experiments the
electron phase deduced from roughly assuming a constant group
velocity of $v_{{\rm g}}\sim 2-5.10^{6}$ cm/sec disagrees with the
single particle picture, and the authors argued this in favor of a
possible breakdown of the single particle picture and the
Landauer-Buttiker conductance formalism.

In the present work, we systematically analyze the electrostatic
edge profile of narrow Hall bar samples using a self-consistent
Thomas-Fermi-Poisson approach (SCTFPA) under QH conditions. In
Sec.~\ref{sec:2}, we introduce a model which incorporates a
constant donor density ala the Chklovskii
geometry~\cite{Chklovskii93:12605}, and secondly
(Sec.\ref{sec:dd}) by a nonuniform distribution of donors. We then
find, in Sec.~\ref{sec:dd} and Sec.~\ref{sec:sec4} the electron
velocity $v_{{\rm y}}$, considering different models, across the
sample in the current direction by \be v_{\rm
{y}}=\frac{1}{\hbar}\cdot\frac{\partial E_{{\rm X,n}}}{\partial
k_{{\rm y}}}, \ee where $E_{{\rm X,n}}$ is the eigenenergy of the
single particle Hamiltonian with $X=\hbar k_{{\rm y}}/eB$,
denoting the center coordinate, and $n$, the Landau Level (LL)
index. Here, $k_{{\rm y}}$ is the conserved electron momentum in
$y$-direction, $e$ is the electron charge and $B$ represents the
strength of the perpendicular magnetic field. We then investigate
the dependence of the electron velocities on the $B$ field
considering two edge state models in Sec.~\ref{sec:sec4}. The
widths of the ISs depending on the sample properties are examined
in Sec.~\ref{sec:sec5}. We observe that the electron transport is
confined within the ISs where the electron velocity decreases with
increasing magnetic field as $B^{-1/2}$. On the other hand, if the
center filling factor, $\nu(0)$, is smaller than its minimum
integer value $2$ (since, we do not resolve the spin degeneracy),
all the current is spread over the sample, suggesting that the
slope of the screened potential should be calculated at the Fermi
level. We close our discussion with a summary section.

\section{\label{sec:2} The Model}
The 2DES, described by the electron number density $n_{\rm
el}(x)$, is considered to be in the $x-y$ plane with a lateral
confinement $|x|<b$ at $z=0$ and assuming translation invariance
in the $y$-direction. The ionized donors also reside in this
plane, with the average number density $n_{0}$ confined into the
interval $|x|<d$, where $d$ is the sample width and $(d-b)$ the
depletion length with $ b < d $. Electrostatic self consistent
solution is then independent of the $y$ coordinate and from the
solution of the Poisson's equation with the appropriate boundary
conditions, i.e. $V(x=\pm d,z=0)=0$, we obtain the Hartree
potential energy of an electron in the plane of the 2DES as
\be V_{{\rm H}}(x)=  \frac{2e^2}{\bar{\kappa}} \int_{-d}^{d} dx'\,
K(x,x')\, n_{{\rm el}}(x'), \label{eq:VHartree} \ee
with $\bar{\kappa}$ being the dielectric constant of the material
and the electrostatic kernel~\cite{Siddiki03:125315} is \be
K(x,x')=\ln
\left|\frac{\sqrt{(d^2-x^2)(d^2-x'^2)}+d^2-x'x}{(x-x')d} \right|
\,. \label{kernel-inplane} \ee The potential energy of an electron
in the confinement region generated by the donors reads:
\be V_{\rm bg}(x)=-E_0 \sqrt{1-(x/d)^2}\,, \quad E_0=2\pi e^2 n_0
d/\bar{\kappa} \, , \label{vbg-inplane} \ee
which can be found from Eq.~(\ref{eq:VHartree}) using the kernel
given in Eq.~(\ref{kernel-inplane}) and replacing $n_{\rm el}(x')$
by $-n_0$. We write the total potential energy of an electron as
$V(x)=V_{\rm H}(x)+ V_{\rm bg}(x)$. The electron number density is
calculated numerically, within the Thomas-Fermi Approximation
(TFA)
\be n_{\rm el}(x)=\int dE\,D(E)\,f([E+V(x)-\mu]/k_{\rm{B}}T)
\label{tfaed} \ee
with $D(E)$ describing the (collision-broadened) Landau density of
states (DOS), $f(\alpha)=(1+e^{\alpha})^{-1}$, the Fermi
distribution function and $\mu$ the electrochemical potential.
Here, $k_{\rm B}$ and $T$ represent the Boltzmann constant and the
electron temperature respectively. We also assume the electron
spin degeneracy. Eq.'s (\ref{eq:VHartree}) and (\ref{tfaed})
complete the self-consistent
scheme~\cite{Guven03:115327,siddiki2004}, which can be solved by a
numerical iteration. For accurate convergence, we first perform
calculations at $T=0$ and $B=0$, then increase $T$ at an elevated
$B$ strength, and reduce the temperature stepwise until the
relevant temperature is achieved. In the next sections, we first
consider two distribution functions for the donor number density,
and investigate the electron velocity dependence on the magnetic
field, temperature and confining potential profile.

\section{Donor Distribution\label{sec:dd}}

During the last decades, several boundary conditions were
considered, ranging from infinite hard-wall
potentials~\cite{Wulf88:4218} to smooth
potentials~\cite{Chklovskii93:12605,Oh97:13519,Siddiki03:125315}
in order to theoretically investigate the 2DES under QH
conditions. For relatively large samples ($d\gtrsim 1$ mm) the
edge effects were considered to be dominated by the localization,
thus the positions of the ISs were mainly predicted by the
disorder potential. On the other hand, for narrow samples ($d<15
\mu$m) the ISs were considered to be formed due to the
electrostatic boundary conditions at the edges. Recently it was
experimentally shown that, the steep potential at the edge of the
sample prohibits the formation of the ISs and the Chklovskii
picture is no longer applicable~\cite{Grayson05:016805}. These
results coincide with an early theoretical calculation, based on
Hartree approximation, given in Ref.~[\onlinecite{Wulf88:4218}],
where the edge potential is taken to be an infinite wall, for
which ISs were not observed. On the other hand, if the edge profile
is smooth, several incompressible regions can be observed
theoretically~
\cite{Siddiki03:125315,Chklovskii93:18060,Oh97:13519} which are
confirmed experimentally~\cite{Wei98:1674}. In the intermediate
case, corresponding to narrow samples, only a single
incompressible edge strip was
reported~\cite{Ahlswede02:165,Ahlswede01:562} which was then
supported by subsequent theoretical
works~\cite{siddiki2004,Siddiki:ijmp}.

In this section we consider narrow samples ($1\mu$m $\lesssim d
\lesssim 5 \mu$m) and vary the donor distribution, to investigate
the widths of the ISs depending on the magnetic field strength.
The selected donor profiles can be realized experimentally either
by the uncontrollable etching processes or by gradually doping the
sample.

\begin{figure}
{\centering
\includegraphics[width=1.\linewidth]{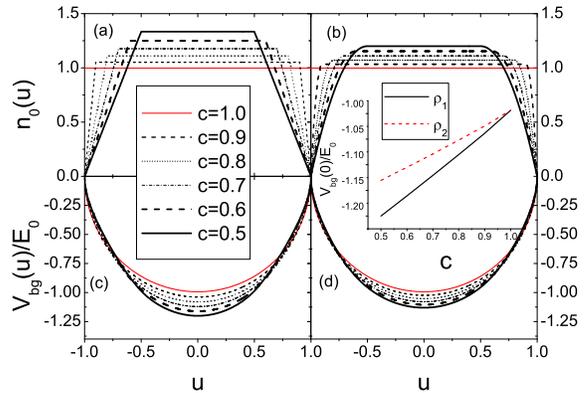}
%
\caption{ \label{fig:fig1} The cross-section of the donor layer
considering (a) $\rho_1(u)$ and (b) $\rho_2(u)$ for various values
of the steepness parameter ($0.5\leq c \leq 1$), together with the
calculated background potential profiles (lower panels, (c) and
(d)). The thin solid line represents a constant donor distribution
($c=1$), whereas thick solid line corresponds to $c=0.5$. The line
code denotes a gradual increase of $c$ with a step of ten percent.
The sample width $d$ and the depletion length, $b$ are fixed and
set to be $3 \mu$m, $300$nm, respectively. In both cases, the
donor number density is kept constant and chosen to be $4\cdot
10^{11}$cm$^{-2}$. The inset depicts the variation of the
background potential calculated at the center of the sample for
$\rho_1$ (thin solid line) and $\rho_2$ (broken line).}}
\end{figure}

In Fig.~\ref{fig:fig1} we show the two selected donor
distributions (upper panel) and the corresponding confinement
potentials (lower panel) generated by \be
\rho_{1}(u)=\left\{%
\begin{array}{ll}
    \frac{[-(u+c)^2+(c-1)^2]n_c}{(c-1)^2}, & \hbox{$-1\leq u<-c$;} \\
    n_c, & \hbox{$-c\leq u \leq c$;} \\
     \frac{[-(u-c)^2+(1-c)^2]n_c}{(c-1)^2}, & \hbox{$c<u\leq1$.} \\
\end{array}%
\right.
\label{donor_dist_1}
\ee
and
\be \rho_{2}(u)=\left\{%
\begin{array}{ll}
    \frac{(u+1)n_c}{(1-c)}, & \hbox{$-1 \leq u<-c$;} \\
    n_c, & \hbox{$-c \leq u \leq c$;} \\
     \frac{(1-u)n_c}{(1-c)}, & \hbox{$c<u\leq 1$.} \\
\end{array}%
\right. \label{donor_dist_2} \ee where $u=x/d$, and $n_c$ is a
constant density preserving the total number of the donors in the
sample. The steepness of the confinement is controlled with the
dimensionless parameter $c$. In the figures, the potential
energies are also normalized with the pinch-off energy ($E_0$) of
the constant donor distribution. To make a connection between the
experimental realization of such donor distributions, we point
that, during the chemical etching in the $z-$ direction the
reaction also takes place in the $x-y$ plane. Hence, the donor
layer is not necessarily etched completely at the edges and a
distribution similar to $\rho_1(u)$ is expected. Meanwhile, during
the growth process of the wafer, donors can be distributed (in a
controlled way) similar to, $\rho_2(u)$. In our calculations in
both cases, we keep the average donor number density constant. It
is clearly seen in Fig.~\ref{fig:fig1} that the steepness profiles
close to the edges of the sample is different for two distribution
functions, whereas  the minima of the confining potentials change
linearly with $c$. As a result, different behaviors can be
identified for the screened potentials (even without exploiting
the magnetic field) arising from the momentum ($q$) dependence of
the Thomas-Fermi dielectric function, i.e.
$\epsilon(q)=1+\frac{2}{a_{\rm{B}}^{*}|q|}$ dominated by
$q=2\pi/a$.

The relation between the screened potential, $V_{\rm {scr}}(q)$
and the external potential $V_{\rm ext}(q)$ is given by, \be
V_{\rm{scr}}(q)=V_{\rm {ext}}(q)/\epsilon(q). \ee Different
steepness values lead to different dominating $q$ regions which
then render different characteristic screening properties. From
the inset of Fig.~\ref{fig:fig1} we conclude that, the steepness
of the potential increases much faster for $\rho_1$, better
simulating the edge profile than the doped profile. This
difference becomes crucial, when a magnetic field is applied and
the ISs are formed at the edges of the sample.

\begin{figure}
{\centering
\includegraphics[width=1.1\linewidth]{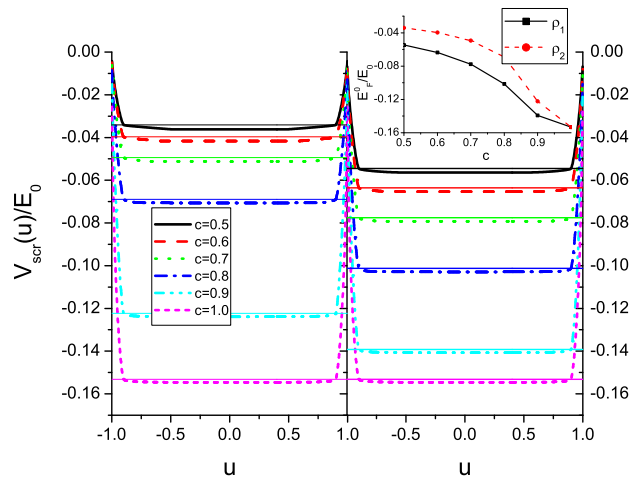}}
%
\caption{ \label{fig:fig2} The screened potentials obtained from
the bare confinement potentials shown in Fig.~\ref{fig:fig1}, at
$B=0$ and $T=0$ (thick lines). Also the Fermi energy for vanishing
field and temperature (thin horizontal lines). The inset depicts
the variation of $E^0_{\rm F}$ versus the steepness considering
$\rho_1$ (solid line) and $\rho_2$ (broken line). }
\end{figure}

\begin{figure}
{\centering
\includegraphics[width=1.1\linewidth]{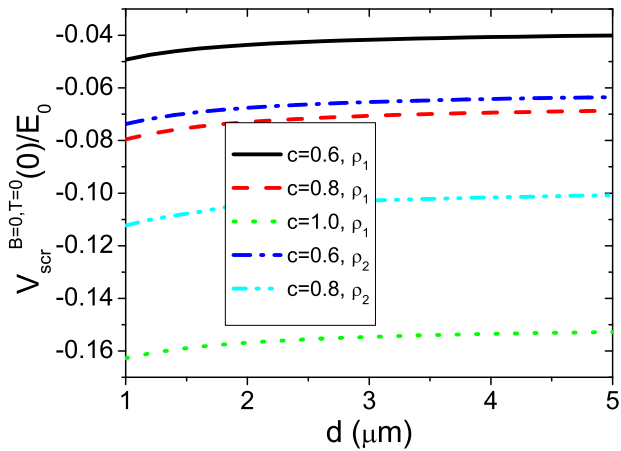}}
%
\caption{The sample width dependence of the screened potential
calculated at the center. The line code depicts the selected
values of $c$ and two distribution functions.\label{fig:fig3}}
\end{figure}

In the next step we consider the effect of the electronic
screening. We calculate the screened potential self-consistently
by solving Eq.'s (\ref{eq:VHartree}) and (\ref{tfaed}) at $T=0$
and $B=0$. In this limit, Eq. (\ref{tfaed}) is reduced to \be
n_{\rm el}(x)=D_0(E^{0}_{\rm F}-V(x))\Theta(E^{0}_{\rm F}-V(x))
\label{tfaedt0}, \ee which then becomes a linear relation between
the potential and the electron distribution within the linear
screening regime. In Fig.~\ref{fig:fig2} the calculated
self-consistent potentials are depicted for the considered donor
distributions together with the variation of the Fermi energy as a
function of $c$. On one hand, the screened potential within the
sample coincides with the $E_{\rm F}^0$ for both models in
(\ref{donor_dist_1}) and (\ref{donor_dist_2}). On the other hand
these quantities differ strongly for both distributions, due to
the non-linear screening pronounced above. Since the constant part
of the donor distribution ($q=0$ component) is strongly screened,
the change of the potential depending on steepness is less
pronounced for $\rho_2(x)$, meanwhile the sharp transition is
observed at $\rho_1(x)$. This implies that more $q$ components
being involved in the screening there. This behavior can be seen
from the slope shown in the inset of Fig.~\ref{fig:fig1}. In other
words, less $q$ components contribute, for $\rho_1(x)$, to the
screened potential in the bulk, whereas, more $q$ components are
involved close to the edge of the sample. Thus the minimum of the
screened potential changes faster than that of $\rho_2(x)$. The
Fermi energies, show a similar behavior, up to a factor, which
indicates that the average number of electrons decreases faster
for $\rho_1(x)$ although the density of the donors is kept
constant.

The effect of the sample width on the potential profile affects
 the variation of the screened potential. In
Fig.~\ref{fig:fig3} the self-consistent potential at the center is
plotted against the (half) sample width for the two donor
distributions in Eq.'s ({\ref{donor_dist_1}) and
(\ref{donor_dist_2}) for the selected steepness parameters. For
large samples ($2d\gtrsim 6 \mu m$), the variation of the central
value of the potential is not sensitive to the steepness, since
the electrons at the bulk can screen perfectly the confinement
potential at the edges. On the other hand, steepness is expectedly
important for narrower samples. This observation clarifies the
dominating role of the edge profile on the electron velocity for
narrow samples and shows that the interaction effects become
important in the Mach-Zehnder type experiments, where the
dimensions of the samples used are usually less than a few
micrometers ($2d<3-4 \mu m$).

\begin{figure}
{\centering
\includegraphics[width=1.1\linewidth]{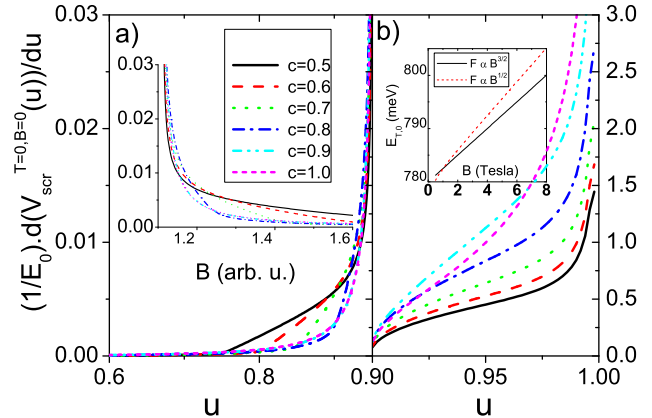}}
%
\caption{ \label{fig:fig4} The numerical derivative of the
screened potentials for different steepness values calculated at
the edge of the sample for inverse parabolic donor distribution,
$\rho_2$. Horizontal axis essentially presents the Fermi level,
i.e. increasing of the $u$ corresponds to increasing of the
average electron density.}
\end{figure}

Before proceeding with the investigation of the effects of the
magnetic field, i.e. considering the effects of the IS in the
presence of nonlinear electronic screening, we briefly discuss the
slope of the self-consistent Hartree potential by only taking into
account the LL quantization. As the magnetic field is changed, the
LLs and their separations are shifted on the energy axis. A
qualitative understanding of the $B$ dependence of the induced
electric field ($F$) within the sample can then be achieved by
analyzing this shift. Then $F(B)$ can be compared with the
experimental results obtained by
Ref.~[\onlinecite{Karmakar05:282}]. Our main argument is, even
without taking the ISs into account, one should be able to observe
the predicted behavior of the (average) electron velocity
($v_{el}$) at the edges of the sample.

We use the derivative of the screened potential with respect to
$u$ at the Fermi level to infer $v_{\rm el}$. This derivative of
$V_{\rm scr}(u)$ is shown in Fig.~\ref{fig:fig4}, for selected
values of $c$ and using $\rho_2(u)$ as the donor distribution. We
observe two characteristic behaviors. In Fig.~\ref{fig:fig4}a, in
the electron dense region ($|u|<0.9$ where $V_{\rm
scr}(u)/E_0<E_{\rm F}/E_0$), the derivative grows in the positive
$x$ direction rather slowly compared to that in the electron
depleted region, as shown in Fig.~\ref{fig:fig4}b. Screening is
strong in the electron dense interior where the total potential is
flat. Approaching the edge of the populated region, the number of
electrons decreases and screening becomes poor. In the depleted
region, the confinement potential is screened very poorly, thus
the variation of the total potential there is large yielding a
larger derivative. For a given Fermi energy, decreasing the
magnetic field corresponds to sweeping the $x$ axis by which the
electron velocity along $y$ can be deduced. In the inset of
Fig.~\ref{fig:fig4}, $v_{\rm el} \propto B^{-1/2}$ behavior is
clearly observed for all $c$ values, however, the exact
quantitative values depend on the steepness of the edge profile.
We observe that, for $c<0.8$ the change in the electron velocity
is much more rapid in this case than the shallow edge profiles,
indicating a strong relation between the edge profile and the
magnetic field dependence of the electron velocity. Combining
$v_{\rm el} \propto B^{-1/2}$ and the calculated slopes
demonstrate that, in the mentioned
experiments~\cite{Karmakar05:282} the confinement is relatively
steep; which was concluded by these authors to be the opposite. In
connection, here we would like to stress another experiment where
a similar geometry reported in
Ref.~[\onlinecite{Grayson05:016805}] was considered. In this work
it is ruled out that, if a negatively charged gate is placed on
the side perpendicular to the 2DES, (in the experimental setup
this gate is an other 2DES, obtained by a cleaved edge overgrowth
technique) creating a sharp potential profile at this edge, no ISs
are observed. Similarly, a side (gate) electrode is used to detect
the SPV signal in the experimental setup of Karmakar
et.al~\cite{Karmakar05:282} and their conclusion contradicts
strongly with the findings of Huber et.
al~\cite{Grayson05:016805}. They also contradict with the velocity
dependence, which we discuss next in more detail now also
including the incompressible regions. It is clear that we will
work with those geometries where the edge profile is neither very
steep (such as an infinite wall or a perpendicular side gate) nor
very shallow so that many incompressible regions can be observed
at a given magnetic field, within Thomas-Fermi approximation.

\begin{figure}
{\centering
\includegraphics[width=1.1\linewidth]{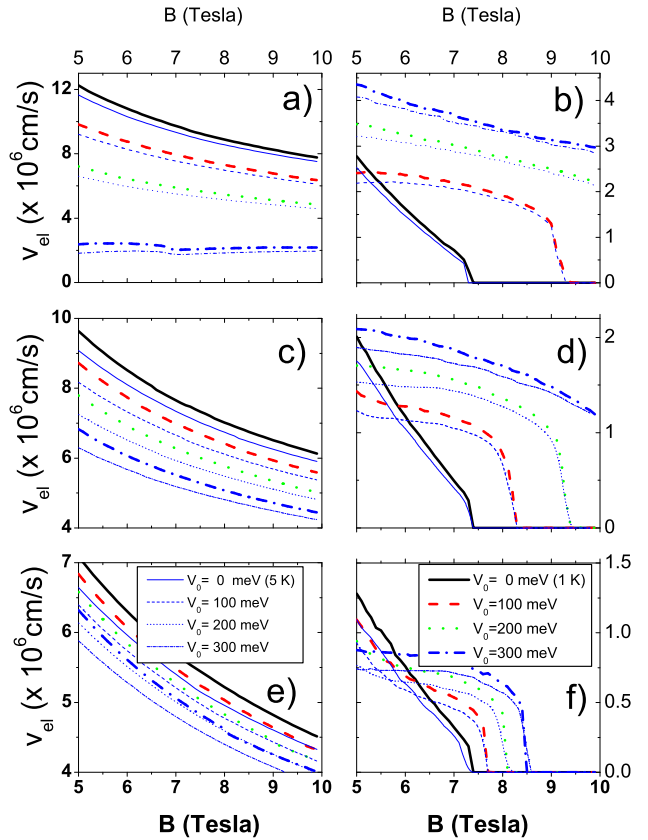}}
%
\caption{The slopes of the screened potential calculated at the
Fermi level (left panel) and within the IS
(right panel), considering characteristic (half) sample widths of
$d=1\mu$m (upper panel), $d=2\mu$m (middle panel) and $d=5\mu$m
(lower panel). The electron temperatures are chosen to be $T=1$K
(thick lines) and $T=5$K (thin lines). \label{fig:fig5} .}
\end{figure}

In the SPV work, the Hamiltonian of the system was given by \be
H=\frac{1}{2m_{e}^{*}}(\mathbf{p}-e\mathbf{A})^2+eFx \ee which
includes a constant electric field ($F$) along the positive $x$
axis pointing to the edge. Here $m_{\rm e}^{*}$ is the effective
electron mass and $\mathbf{p}$ and $\mathbf{A}$ are the canonical
electron momentum and the vector potential respectively. Using the
Landau gauge, the energy dispersion is found to be~\cite{Davies}
\be
E_{n,X}=E_{\rm g}+(n+1/2)\hbar\omega_c && -(F/B)(X/l_{\rm b}^{2}) \nn \\
&&-(m_{\rm e}^{*}/2)(F/B)^2, \label{energyk} \ee where $E_{\rm g}$
is the energy band gap and $l_{\rm b}=\sqrt{ \hbar/m\omega_c}$ the
magnetic length. These authors concluded that, in order to obtain
the $B=0$ value and also to match the experimental results (see
Fig.4 of Ref.[~\onlinecite{Karmakar05:282})], one should assume
that $F\propto B^{3/2}$. First of all, one remark is that, in the
$B=0$ limit energy dispersion given in equation (\ref{energyk})
becomes meaningless. Secondly, in the limit of high magnetic
field, assuming $F\propto B^{3/2}$ or $F\propto B^{1/2}$
essentially leads to similar linear behavior at the measured $B$
values, as shown in the inset of Fig.~\ref{fig:fig4}. In the SPV
experiments no low field ($B\lesssim 1$ T) measurement were
performed, therefore  we conclude that their conclusion about
$F\propto B^{3/2}$ in Ref.~[\onlinecite{Karmakar05:282}] is not
unique. Moreover our calculations ascertain that, even in the
absence of ISs, the electron velocity is an inverse square root
function of the magnetic field, namely $v_{\rm el}\propto
B^{-1/2}$, and as a consequence, $F\propto B^{1/2}$.

Our simple self-consistent calculations, assuming that the effects
of ISs are negligible, agree qualitatively well with the
experimental findings. We also point that the functional form of
the electric field and the interpretation of the steepness of the
potential strongly differ from Ref.~[\onlinecite{Karmakar05:282}].
First of all, it is experimentally~\cite{Grayson05:016805} and
theoretically~\cite{Wulf88:4218,Siddiki03:125315} shown that in
the presence of a side gate, perpendicular to the 2DES,
(simulating a hard wall potential or surface charges) the
potential at the edge is steep. Secondly, the proclaimed $B$
dependence of the electric field at the edge is not unique and we
claim that $F\propto B^{1/2}$.

The discussion above should also be reconsidered in the presence
of IS. In the next section we do that by examining both the
potential slope at the Fermi level and at the position of the ISs to
obtain a more realistic comparison between our theory and the
recent Mach-Zehnder interferometry
experiments~\cite{Heiblum03:415,Neder06:016804}.
\begin{figure}
{\centering
\includegraphics[width=1.1\linewidth]{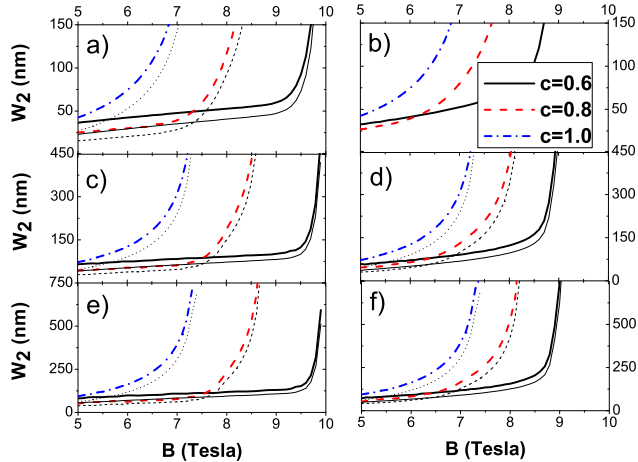}}
%
\caption{The sample width dependence of the IS thickness for
$\nu(x)=2$. Calculations are performed at $1$K (thick lines) and
$5$K (thin lines) for three characteristic steepness values
considering etched (left panel) and functionally doped (right
panel) samples. Widths of the samples are selected to be $d=1
\mu$m (top) $d=3\mu$m (middle) and $d=5\mu$m (bottom), whereas the
electron depleted strips are fixed to be the ten percent of $d$.
\label{fig:fig6} }
\end{figure}

\section{Comparison of the E-field at different edge state models\label{sec:sec4}}
In principle, the electron velocity at the edge of the 2DES or the
electric field at the depleted region is not directly measured in
the SPV experiments, instead the slope of the potential profile is
investigated as a function of the magnetic field deduced by the
energy dependence of the SPV signal. On the other hand, in our
calculations we explicitly obtain the self-consistent potential
and by taking the derivative of the energy dispersion we can
directly calculate the electron velocity. In the previous section,
by making use of the Thomas-Fermi approximation, we obtained the
screened potential and claimed that the center coordinate
dependent dispersion is given by, \be E_{n}(X)=E_{n}+V_{\rm
scr}^{T=0,B=0}(X). \ee The next step is to calculate the full
$V_{\rm scr}^{T\ne 0, B\ne 0}(X)$ and to investigate its slope as
a function of $B, T, c, d$ as well as the long-range part of the
disorder potential.

As a standard
technique~\cite{Siddiki03:125315,Siddiki:ijmp,siddikikraus:05,Siddiki02:Oxford},
we simulate the potential fluctuations generated by the disorder
by imposing a modulation potential~\cite{Siddiki02:Oxford} of the
type \be V_{\rm m}(x)=V_{0}\cos(k_{\lambda}x), {\rm with} \quad
k_{\lambda}=(\lambda+1/2)\pi/d \ee as an additive contribution to
the confinement potential. Here $V_0$ is the modulation amplitude
and $\lambda$ is an integer to preserve the boundary conditions.

In Fig.~\ref{fig:fig5}, we show the numerical derivative of
$V_{\rm scr}(x)$ at the chemical potential. Note that at $T \ne 0$
the Fermi energy is no longer equal to the chemical potential, and
it has to be calculated for the given set of physical parameters.
Here we consider $\lambda=5$. The amplitude of the modulation is
set such that, after screening, the potential variation is at the
order of $\% 5-25$ of $E_{\rm F}^0$. As a rough estimate we find
that $V_0$ reduces by three orders of magnitude due to the
dielectric screening, for GaAs, $\kappa =12.4$, and electronic
screening ($\epsilon(q)\sim 41$, see e.g. the expression given in
the caption of Fig.2 of Ref.~[\onlinecite{Siddiki03:125315}] and
the related text). Regardless of the variation in $d$, $T$ and
$V_0$, the slope of the screened potential obtained at the
chemical potential exhibits the previously observed $F\propto
B^{-1/2}$ form; except the case where the modulation is so strong
that slope remains unaffected at $d=1\mu$m with $V_0=300$ meV. The
hint to understand this exception is found in
Fig.~\ref{fig:fig5}b, where we show the slope calculated inside
the IS at the position corresponding to $\nu(x)=2$. We see that
the derivative of the potential also behaves similar to the one
obtained at the chemical potential, namely the inverse square root
form, which indicates that the ISs are considerably narrow. Also
from the density profile calculated (not shown here, however, the
results of a similar calculation can be found in
Ref.~[\onlinecite{Siddiki02:Oxford}]), we see that, due to the
strong modulation, the outermost IS is narrow and it's effect is
negligible, thus the slope remains almost insensitive to the
change in the $B$ field on this scale. An interesting comparison
concerning the sample widths reveals that the narrower the sample
is, the stronger the slope. Hence in the Mach-Zehnder experiments
and also considering the fact that the measurements are performed
at an intermediate magnetic field strength ($B\sim 2.5-4.5$T) and
narrow samples ($d\sim 1\mu m$), the assumption of a constant
velocity independent of $B$ is not realistic. We observe that, the
disorder potential does not effect this general behavior as long
as the dominating scattering processes come from the edges of the
sample. Introducing disorder obviously results in density
fluctuations, which can be screened by the 2DES if the system is
compressible (far from integer filling factors) and the conclusion
is the opposite if the Landau levels are fully occupied. We
consider a situation such that the magnetic field is tuned to an
interval, where the average filling factor becomes close to an
(even) integer. In this situation, a large IS is formed at the
bulk (without modulation), split into several ribbons (as observed
in Fig. 2 of Ref.~[\onlinecite{Siddiki:ijmp}]) and the effect of
these incompressible ribbons on the slope at the edge is marginal.
This is seen in the left panel of Fig.~\ref{fig:fig5}, where we
examine the behavior of the derivative comparing $V_0=0$ and
$V_0\neq 0$. In the unmodulated case the slope drops linearly with
increasing $B$, until a large IS is formed at the bulk (e.g. in
Fig.~\ref{fig:fig5}f $B\sim 7.3$T). The wide strip disappears when
the magnetic field strength is strong enough so that the Fermi
level is pinned to the lowest Landau level, $B\sim 7.45$T. For the
modulated case the derivative decreases also linearly, with a
smaller slope. However, this linear region is larger compared to
the unmodulated case, e.g. in figure \ref{fig:fig5}d up to $B\sim
7.8$T for $V_0=100$ meV and $B\sim 9.0$ T for $V_0=200$ meV.
Depending strongly on the modulation amplitude, the rapid decay of
the slope due to the formation of a large bulk IS, is observed in
a relatively narrow $B$ interval. The ``linear slope regime'' is
observed for all considered sample widths; however, for narrow
samples the $B$ interval is larger for higher modulation
amplitudes. This indicates that for high mobility samples, where
the long-range part of the disorder potential is well
screened~\cite{Siddiki:ijmp}, the linear regime will be observed
in a narrow $B$ interval. From the above discussion we conclude
that, the electron velocity on the ISs presumes a linear $B$ field
dependence. At this point we find it useful to make a connection
between our results and the Mach-Zehnder type samples. These
samples have intermediate mobility and are relatively narrow. We
have shown that the electron velocity calculated at the chemical
potential decreases like an inverse square root of the $B$ field
and the assumption of constant velocity is not applicable.
Whereas, if the current is carried by the ISs, assuming a constant
$v_{el}$ in the magnetic field interval where the interference
pattern is observed is still irrelevant. Recently it has been
shown theoretically that~\cite{SiddikiMarquardt}, within the
screening picture of integer quantized Hall effect, the
interference can be observed only in a narrow magnetic field
interval within the plateau regime at high mobilities. The
boundaries to observe interference pattern is estimated such that
two separated ISs should be formed (similar to $B<8$T of figure
\ref{fig:fig6}f) which are larger than the Fermi wave length
($B>6$T). This interval coincides with the linear velocity regime
shown above. Therefore we support the idea~\cite{Neder06:016804},
that the phase of the electron calculated within the single
particle picture should also be reconsidered form the interaction
point of view as presented in this work. So far we have examined
the magnetic field dependence of the slope of the screened
potential at the Fermi level and within the ISs. We have found
that, depending on the $B$ strength, the electron velocity
exhibits different behaviors depending on where the slope is
calculated. If it is assumed that the current flows from the
Landauer-B\"uttiker type edge states the velocity takes the form
$B^{-1/2}$. Whereas, if the current is carried by the ISs, the
velocity drops linearly in the case of two separate ISs and is
highly non-linear in the presence of a large IS in the bulk. Next,
we discuss the extend of the ISs depending on the magnetic field
and steepness of the confinement potential considering different
sample widths and temperatures.

\section{The formation of IS\label{sec:sec5}}
The long-standing question of ``where the current flows ?'' in the
quantum Hall bar systems has been addressed in many different
theoretical
works~\cite{Guven03:115327,siddiki2004,Buettiker86:1761,Cooper93:4530,Chang90:871,Han97:1926}.
In a recent model using a local version of the Ohm's law, it was
shown that the external current is confined in the ISs where the
longitudinal resistivity vanishes, i.e. $\rho_\ell(x)=0$. This
novel approach brought a quantitative explanation to many
interesting aspects of the integer quantized Hall effect, among
which are the high reproducibility of the very accurate quantized
Hall plateaus, the transition between the zero states and the
description of the local current distribution. This model is based
on the formation (and disappearance) of the ISs, and now we
concentrate on their widths taking into account different edge
profiles and sample widths.

In Fig.~\ref{fig:fig6}, we plot the widths of the ISs ($W_2$, for
local filling factor 2) against the magnetic field strength
considering different sample properties. For the constant donor
distribution ($c=1$) we see that the sample width has no influence
on the $B$ dependence of the width of the strips. Whereas $W_2$
increases by increasing the sample width, as expected. Note that,
since the variation of the self-consistent potential at the IS is
$\hbar \omega_c$, the slope is calculated simply by dividing this
variation by $W_2$. As a direct consequence, the slope becomes
small when $W_2$ becomes large. At the first glance, for
intermediate steepness ($c=0.8$) the functional form of the
inhomogeneous donor distribution, i.e. $\rho_1(x)$ or $\rho_2(x)$,
has no influence on $W_2$. However, the linear velocity regime is
much more extended for the etched samples than of the doped ones
for relatively large sample widths ($d\gtrsim 3\mu$m). This
feature is more pronounced for $c=0.6$, i.e. for the steeper edge
profile, and the linear (velocity) regime is observed in a larger
magnetic field interval compared to other steepness parameters.
The slope of the linear regime is smaller for the etched sample
and a smoother transition to the non-linear regime is observed for
the doped edge profile, whereas the functional form of the donor
distribution seems to show no important difference for different
sample widths.

As a final remark on the IS widths, we would like to recall the
findings of G\"uven and Gerhardts~\cite{Guven03:115327} where the
high current regime was also investigated. It was shown that, a
large imposed current leads to a broadening of the ISs on one side
of the sample, hence a change in the slope, which was also
supported by the experiments~\cite{Ahlswede02:165}. This result
shows that there is a relation between the amplitude of the
imposed current and the average electron velocity inside the ISs.
We believe that, the investigation of the out-of-the-linear
response regime will improve our understanding of the Mach-Zehnder
type of interferometer experiment. Our preliminary results show
that, the widths of the ISs increase linearly by increasing the
amplitude of the applied current.

\section{Summary}
In summary, we have calculated the slope of the self-consistent
potential, within the Thomas-Fermi-Poisson theory of screening. We
considered two different pictures of edge states, namely the
single particle and the incompressible states, to obtain electron
velocities in the presence of a strong perpendicular magnetic
field. We have systematically investigated the effect of the
sample properties such as the sample width, edge profile and
disorder potential on the electron velocities.

We first obtained a functional form of the $v_{\rm el}$ and the
electric field depending on the magnetic field strength, without
taking into account the formation of the ISs, and considering only
the Landau quantization. It is shown that the interpretation of
the SPV experiments~\cite{Karmakar05:282} strongly contradicts
with our results and also with other
experiments~\cite{Grayson05:016805}. We found that the slope of
the self-consistent potential changes as $\sim B^{-1/2}$, whereas
the electric field at the edge behaves as $F\propto B^{1/2}$. We
also concluded that assuming a constant $v_{\rm el}$ may lead to
discrepancies in analyzing the results of Mach-Zehnder
interferometer~\cite{Heiblum03:415,Neder06:016804} type
experiments.

Secondly, by evaluating the full self-consistent potential, we
were able to obtain the electron velocities at the Fermi level and
within the ISs. We found that, the full self-consistent results
coincide with our semi-consistent findings pointing to the inverse
square root dependence of $v_{\rm el}$ within the single particle
picture. The slope of the fully screened potential calculated at
the ISs, however, exhibits two different regimes of magnetic
field. These two regimes are identified by the dependence of the
electron velocity on the magnetic field which is linear in one
regime and non-linear in the other.

Our results indicate that, in narrow Hall bar geometries with
intermediate mobilities, the edge profile becomes very important
in determining the electron velocity for both the
Landauer-B\"uttiker or the IS type edge states. It appears to us
that re-examining the results of Mach-Zehnder
interferometer~\cite{Heiblum03:415,Neder06:016804} experiments
from self-consistent point of view will thus be helpful to
understand the underlying physics of the obtained interference
patterns.
\newline
\textbf{Acknowledgement}

We would like to thank R. R. Gerhardts for his support and
fruitful lectures on screening theory, which in fact enabled us to
understand the basics. This work was financially supported by SFB
631, T\"{U}B\.{I}TAK grant 105T110 and Trakya University research
fund under project numbers T\"{U}BAP-739-754-759. The authors are
also grateful to the Institute of Theoretical and Applied
Physics/Marmaris/Turkey for partial support where part of the work
was carried.

\end{document}